\documentclass{emulateapj}
\usepackage{epsfig}
\usepackage{apjfonts}

\newcommand{\xmm}{{\it XMM-Newton}}

\newcommand{\gs}{GS\,1826--24}

\begin{document}

\def\spose#1{\hbox to 0pt{#1\hss}}
\def\laeq{\mathrel{\spose{\lower 3pt\hbox{$\mathchar"218$}}
     \raise 2.0pt\hbox{$\mathchar"13C$}}}
\def\gaeq{\mathrel{\spose{\lower 3pt\hbox{$\mathchar"218$}}
     \raise 2.0pt\hbox{$\mathchar"13E$}}}

\slugcomment{Accepted for publication in ApJL}

\title{Non-Detection of Gravitationally Redshifted Absorption Lines in
the X-ray Burst Spectra of GS\,1826--24} 

\author{Albert~K.~H.~Kong\altaffilmark{1,2}, 
Jon~M.~Miller\altaffilmark{3},
Mariano Mendez\altaffilmark{4}, Jean Cottam\altaffilmark{5}, 
Walter~H.~G.~Lewin\altaffilmark{1}, Frederik Paerels\altaffilmark{6},
Erik~Kuulkers\altaffilmark{7}, Rudy
Wijnands\altaffilmark{8}, and Michiel~van~der~Klis\altaffilmark{8}  
} 

\altaffiltext{1}{Kavli Institute for Astrophysics and Space Research,
Massachusetts Institute of Technology, 77
Massachusetts Avenue, Cambridge, MA 02139}
\altaffiltext{2}{Department of Physics and Institute of Astronomy,
National Tsing Hua University, Hsinchu, Taiwan; akong@phys.nthu.edu.tw}
\altaffiltext{3}{Department of Astronomy, University of Michigan, Ann 
Arbor, MI 48109}
\altaffiltext{4}{SRON, Netherlands Institute for Space Research,
  Sorbonnelaan 2, 3584 CA Utrecht, Netherlands}
\altaffiltext{5}{Exploration of the Universe Division, NASA Goddard
  Space Flight Center, Greenbelt, MD 20771} 
\altaffiltext{6}{Columbia Astrophysics Laboratory, Columbia
  University, 550 West 120th Street, New York, NY 10027}
\altaffiltext{7}{ISOC, ESA/ESAC, Urb. Villafranca del Castillo, PO Box
  50727, 28080 Madrid, Spain} 
\altaffiltext{8}{Astronomical Institute ``Anton Pannekoek,''
  University of Amsterdam, Kruislaan, Amsterdam, Netherlands}

\begin{abstract}
During a 200 ks observation with the \xmm\ Reflection Grating
Spectrometer, we detected 16 type-I X-ray bursts from \gs. We combined
the burst spectra in an attempt to measure the gravitational redshifts
from the surface of the neutron star. We divided the composite \gs\ burst
spectrum into three groups based on the blackbody temperature during
the bursts. The spectra do not
show any obvious discrete absorption lines. We compare our
observations with those of 
EXO\,0748--676.
\end{abstract}


\section{Introduction}
The physical properties of neutron stars can be described by an
equation of state. For a given equation of state one can obtain a
unique relation between mass and radius. Observational constraints on
the mass-radius relation of neutron stars can be estimated by a number
of methods. One possible way is to measure spectral lines in X-ray
burst spectra (Lewin 1993). Absorption features
at 4.1 keV (Waki et al. 1984; Nakamura, Inoue, \& Tanaka 1988; Maginer et
al. 1989) and 5.7 keV (Waki et al. 1984) were reported in X-ray burst
spectra with {\it Tenma} and {\it EXOSAT}. However, they were never
confirmed with more sensitive instruments.

During a 335 ks observation made in the calibration phase of \xmm,
28 type-I X-ray bursts were detected from EXO\,0748--676. 
Cottam, Paerels \& Mendez (2002) reported features in the
   bursts spectra of this source which they interpreted as
   gravitationally redshifted absorption lines of Fe XXVI
   during the early phase of the bursts, and Fe XXV and
   perhaps O VIII during the late phase. The lines would then
   have all a gravitational redshift of $z=0.35$, which would
   correspond to a neutron star in the mass range of 1.4--1.8
   $M_\odot$ and in the radius range of 9--12 km (see also
   van Paradijs \& Lewin 1987; {\"O}zel 2006). If confirmed, these detections might rule out
   soft equations of state for neutron star matter ({\"O}zel 2006).
   This is potentially very exciting, however, it still needs
   to be confirmed that these features in EXO\,0748--676 are
   gravitational redshifted absorption lines, either by
   observing it in an independent data set of the same source or
   observing similar features in the burst spectra of other
   systems.
Before the above results from EXO\,0748--676 were known,
 \xmm\ observations of the X-ray burster \gs\ were requested in
 October 2001 (PI: Lewin) to search for
gravitationally redshifted lines in X-ray burst spectra.

\gs\ is an ideal source for studying thermonuclear X-ray bursts because of its
bright, long, and regular bursts (Galloway et al. 2004; Kong et
al. 2000). The bursts have recurrence times between $\sim3.5$ hr and
$\sim5.7$ hr
(Cornelisse et al. 2003; Galloway et al. 2004). We report here on the
results of our 200 ks \xmm\ observations.

\section{Observations and Data Reduction}

\gs\ was observed with \xmm\ two times on 2003 April 6--8 (108 ks) and April
8--9 (92 ks) during which a total of 16 type-I X-ray bursts were
observed with a recurrence time of $\sim3.1$ hr. Both
the European Photon Imaging Camera (EPIC) and the 
Reflection Grating Spectrometer (RGS) were turned on; the Optical
Monitor was turned off during the observations. In this 
paper, we present the high resolution X-ray 
spectra from the RGS. We used the EPIC data only for identifying bursts and
examining the burst profiles. The RGS covers the energy range from 5 to 35
{\AA} (0.35--2.5 keV) with a resolving power of $\sim400$ at 15
{\AA}. The EPIC consists
of three detectors (one pn camera and two MOS cameras) sensitive
between 0.2 and 15 keV. In this
analysis, we only
used data from the pn detector (with the thin filter and in timing
mode). The data were 
processed with the \xmm\ Science Analysis System (SAS) version 7.0. We
performed spectral analysis by using XSPEC v11. 

\section{Results}
\subsection{Burst Profiles and Broadband Spectra}
We plot the burst profiles of the 16 bursts in Figure 1; they
are very similar for all bursts. As during all previous
observations with other instruments (Kong et al. 2000; Galloway et al. 2004), the bursts show a 
fast rise, followed by an exponential decay; typical burst durations were
$\sim 200$ s. Using the EPIC-pn data, we
extracted the time-resolved 0.3--10 keV spectra with 5-s
time resolution in the rising phase and 10-s resolution during
decay. We chose a 2000-s section of data prior to the
bursts which we used as our ``background'' for spectral fits to the
individual spectra of the bursts. The response matrices were
created with SAS together with the latest calibration products. 
The net burst--background
emission was well fitted with a simple blackbody model. The burst and temperature 
profiles are plotted in Figure 2.

\begin{figure}[t]
\includegraphics[width=2.5in,angle=270]{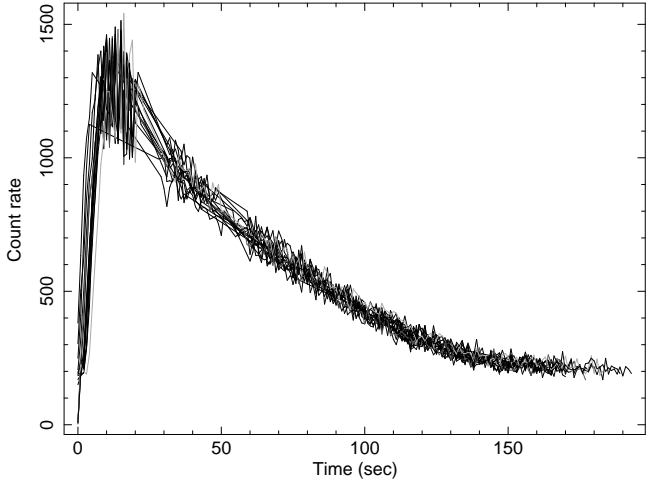}
\caption{Lightcurves of all X-ray bursts detected with the \xmm\ pn
  detector. The missig data during decay are due to telemetry deadtime. All bursts have a similar fast rise followed by an
  exponential decay; burst durations were $\sim 200$ s.
  }
\end{figure}

\begin{figure}[t]
\includegraphics[width=3.5in]{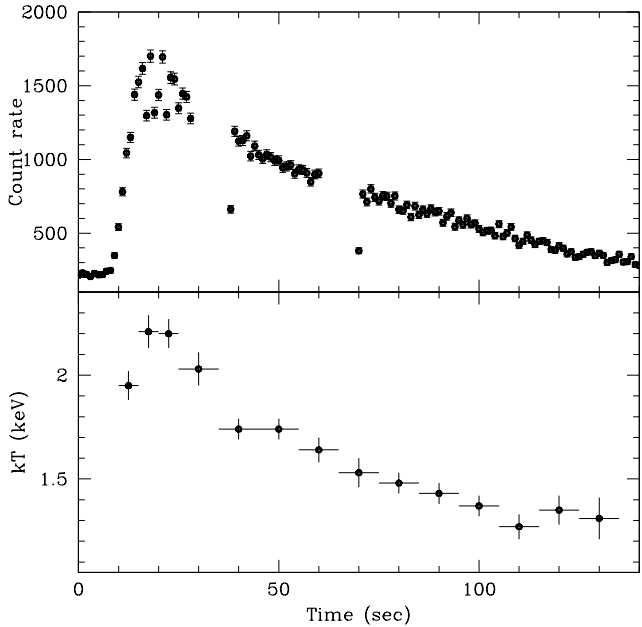}
\caption{A representative burst profile (Upper panel) and temperature
  profile (Lower panel) from \gs\ as observed with the EPIC-pn. The time
  resolution for the lightcurve is 1 s. Drops in count rate are due to
  telemetry deadtime. The rising and decay phase of the
  temperature profile has a time resolution of 5 s and 10 s,
  respectively. The missing data do not show in the temperature
  profile because the time resolution of the two plots is different.}
\end{figure}

\begin{figure}[t]
\includegraphics[width=2.5in,angle=270]{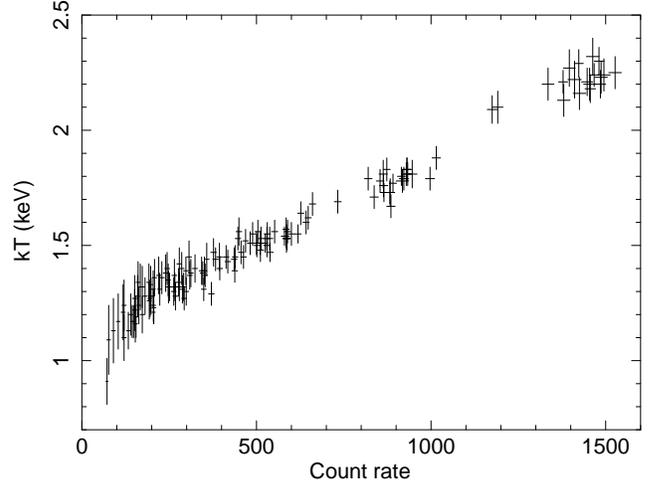}
\caption{Correlation between the blackbody temperatures and count
  rates of all bursts detected with EPIC-pn.}
\end{figure}

\subsection{RGS Burst Spectra}

We used the EPIC-pn light curve as a guide to extract the first-order
RGS spectra for each burst.
Since the spectral properties are changing
during the bursts (see Fig. 2), we studied the RGS spectra separately according
to the blackbody temperatures as derived from the EPIC-pn data. In Figure 3,
we show the correlation between the blackbody temperatures and count
rates of all bursts. A strong positive correlation is found
between the two quantities indicating that the properties are very
similar for all bursts. Using this correlation as a reference, we
divided the bursts into three phases: $kT <1.5$ keV, 1.5 keV $< kT< 2$
keV, and $kT > 2$ keV.  We note that the division we used here is
different from that in the study of EXO\,0748--676 (Cottam et al. 2002)
in which the bursts were divided into ``early'' and ``late'' phases.

For each observation, and for each RGS camera, we extracted three
separate first-order spectra of the bursts according to their blackbody
temperature as determined by the EPIC-pn count rate (Fig. 3). For each
RGS camera we used the same response matrix for each observation.
Finally, we created background spectra using spatially offset regions.
We then used the task {\it rgscombine} in the SAS software to combine the
three RGS spectra of the two observations separately for each RGS
camera. We ended up with 6 spectra in total, one for each interval of the
blackbody temperature of the bursts in each of the two RGS cameras. 
Since all bursts are statistically indistinguishable, for each
temperature interval we combined the spectra 
of both RGS cameras using the SAS command {\it rgsfluxer}. In Figure 4 we
plot the background-subtracted spectra for the three phases of the
average burst. 
Above 27 {\AA} the flux drops significantly due to the effect of the
interstellar absorption.

We rebinned the spectra with at least 20 counts per spectral bin, and
used $\chi^2$ statistics to find the best-fitting parameters. All three
spectra can be adequately fitted with an absorbed blackbody
model with a reduced $\chi^2$ of $\sim 1$. There is no evidence for absorption 
edges apart from those due to the ISM (oxygen and neon are the most
prominent). Any real absorption line must be seen in both RGS spectra
where there is simultaneous wavelength coverage.  The only possible
line significant at the $3\sigma$ level of confidence or higher is Ne
X Ly-$\alpha$ (theoretical wavelength: 12.1339\AA), though it falls in a
range covered by only one RGS spectrum.  Fits with a Gaussian measure
a wavelength of $12.16^{+0.02}_{-0.06}$\AA~ and an equivalent width of
$80^{+30}_{-20}$~m\AA~ ($1\sigma$ errors).  The FWHM is consistent
with zero, indicating the line is not resolved.  Via an F test, the
line is significant at the $3.4\sigma$ level of confidence.  The line
cannot be tied to the stellar surface or even clearly tied to a wind
as its observed wavelength is consistent with zero shift, though Ne X
is seen in some disk winds (Miller et al.\ 2006).

\begin{figure}[t]
\center{
\includegraphics[width=3.5in]{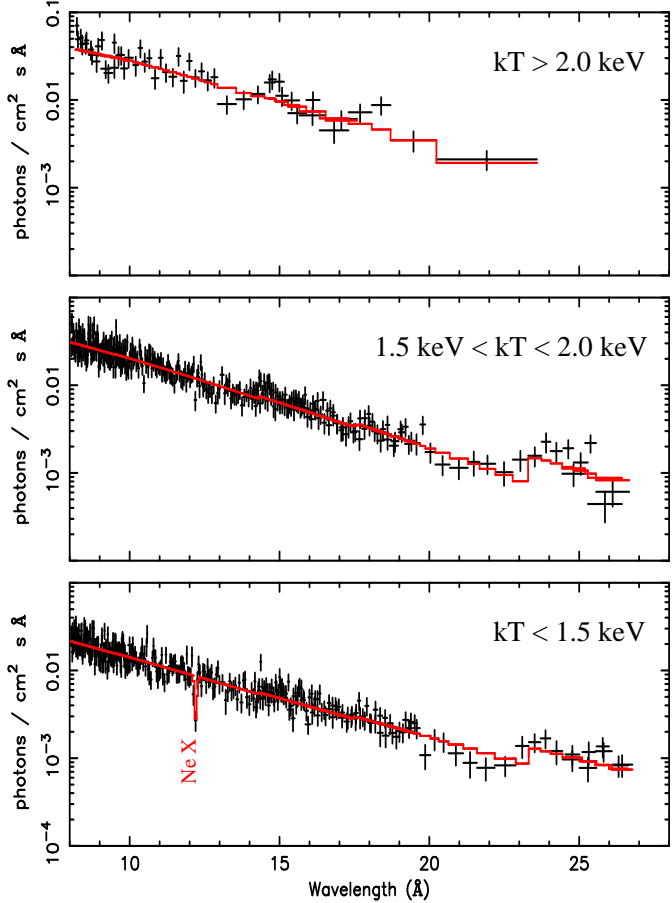}
}
\caption{\xmm\ RGS average background subtracted spectra for 16 X-ray
  bursts from \gs. The three temperature phases are defined according
  to the spectral fits with the EPIC-pn data. An absorbed blackbody
  model is superimposed in red. The Ne X line at 12.1 {\AA} (with
  $3.4\sigma$ confidence) is included in the low-temperature spectrum.
}
\end{figure}

To compare our analysis with EXO\,0748--676, we also 
fitted the three burst spectra with a
circumstellar model similar to that used by Cottam et al (2002).  We
used a continuum model consisting of power-law emission with neutral
interstellar absorption and then added in absorption models for each
of the following ions: N VI, N VII, O VII, O VIII, Ne IX, and Ne X.
We do not intend to establish a definitive spectral fit, but rather to
determine which features might be circumstellar in origin.  We
therefore required that all ions be generated at the same plasma
temperature and redshift, but we allowed the ion abundances to
float. This provided an estimate of the circumstellar contributions
to the spectra, but it does not necessarily provide a physically
realistic circumstellar model for the source.  This fitting process
provides a reasonable estimate for the lower and medium-temperature
spectra; the high-temperature spectrum has a low signal-to-noise ratio
and is difficult to reproduce with the Cottam et al. (2002) model.  
After fitting the circumstellar contributions to the observed burst
spectra, we searched for residual spectral features that might be
generated in the neutron star photosphere.  We do not find any obvious
neutron star photospheric absorption features in these spectra,
consistent with the simple absorbed blackbody model.

In order to obtain upper limits in the equivalent widths of our data in the lines
reported in Cottam et al. (2002), we assume a Gaussian function
centered at between 10--14 {\AA} with a width of $0.1-0.3$ {\AA}. We
note that the width of the lines reported in EXO\,0748--676 is
about 0.11 {\AA} (from unpublished results in Cottam et al. 2002).
Using
these assumptions as well as the low-temperature spectrum, we can set
a 90\% confidence limit to the equivalent width of 0.24
  {\AA} for any absorption line between 10 and 14 {\AA}.  

\section{Discussion}

We observed \gs\ with the \xmm\ RGS and
detected 16 type-I X-ray bursts. 
We note that we are not less sensitive than the results by Cottam et
al. (2002). The reported features from EXO\,0748--676 are from composite
spectra of 28 bursts while we only
detected 16 bursts. However, the total number of counts in the spectra is
comparable because of the longer burst durations in GS 1826-24;
the burst peak intensities were similar.
By adding all the burst spectra,
the high resolution \xmm\ RGS spectra can be fitted with a simple
absorbed blackbody model and show no obvious
absorption features (see Figure 4). 
We do not see in \gs\ features similar to the ones seen by Cottam et
al. (2002) in EXO\,0748--676 which they interpreted as the Fe
XXVI and XXV n=2-3 transitions with a redshift of $z=0.35$. 
For comparison, the equivalent widths of Fe XXV and Fe XXVI lines
reported in Cottam et al. (2002) are 0.18
{\AA} and 0.13 {\AA}, respectively while we obtained a 90\% upper limit
of 0.24 {\AA} for any absorption line between 10--14 {\AA} with a
width of 0.1--0.3 {\AA}.  However,
if the line width is similar to EXO\,0748--676 at 0.11 {\AA}, the 90\% upper limit of the
equivalent width becomes 0.13 {\AA}. We note that follow-up
observations of EXO\,0748--676 have failed to again reveal absorption
lines from the stellar surface (Cottam et al.\ 2007). 

If the features initially seen in EXO\,0748--676 are real and are 
due to absorption lines redshifted by the gravity of
the neutron star, then it is important to determine why we
do not see similar phenomena in our burst spectra from \gs.
Absorption features in burst spectra are very sensitive to the
accretion rate, temperatures, density, and the rotation frequency
of the neutron star (Bildsten, Chang \& Paerels 2003; Chang,
Bildsten \& Wasserman 2005; Chang et al. 2006). For instance,
the different absorption features during the early phase and late
phase of the burst spectra seen from EXO\,0748--676 may be due
to a change in ionization balance of iron in the photosphere. In
the case of EXO\,0748--676, the EPIC-pn burst spectra show a
peak temperature of about 1.8 keV and the temperature drops
to 1.5 keV during burst decay. For \gs\, the peak
temperature of the bursts is about 2.3 keV (see Figure 2). During
decay, the temperature drops to about 1.3 keV. The peak
temperature of bursts from \gs\ is higher than that from
EXO\,0748--676. This may explain why we do not see any
absorption features in the spectrum of burst data in excess of
2 keV. However, the burst temperatures during our low- and
medium-temperature spectra are similar to those observed during
the decay and peak of the bursts from EXO\,0748--676. We
might therefore expect to see absorption features in our spectra.

Rotational broadening may explain the absence of absorption
features in \gs\ burst spectra as it will modify the
line profile ({\"O}zel \& Psaltis 2003). 
Chang et al. (2005) calculated
that for an edge-on neutron star with a rotation rate $> 200$
Hz, the absorption features would become undetectable. 
Thompson et al. (2005) reported a possible
burst oscillation in \gs\ at $\sim 611$ Hz, which, if proven to be true,
would  make the absorption lines less likely to be detected.
EXO\,0748--676, however, is a
very slow rotator with a rotation rate of 44.7 Hz (Villarreal \&
Strohmayer 2004). More recently, Chang et al. (2006) fitted the
line profiles of EXO\,0748--676 with a theoretical model including
the effect of rotational broadening; they reported a gravitational
redshift of $z = 0.345^{+0.005}_{-0.008}$ (95\% confidence). However,
they showed that rapid rotation does not necessarily make spectral
lines undetectable. If a rapidly rotating neutron star is seen
face-on, the spectral lines will be narrower. EXO\,0748--676 is
an edge-on source while \gs\ has an inclination angle
$< 70^{\circ}$ (Homer, Charles, \& O'Donoghue 1998). A theoretical
study shows that narrow lines can be detected even at spin frequencies
as high as 600 Hz for emission near the rotation axis
(Bhattacharyya et al. 2006). It is therefore uncertain whether
a possible high spin frequency of the neutron star could be the
cause of the non-detection of spectral lines in \gs\ since
most accreting neutron stars for which the spin has been determined
have a spin frequency $< 600$ Hz (e.g., van der Klis 2006).

The chemical composition of the atmosphere of the neutron
star is another factor that can influence the line profiles (Chang
et al. 2005). A more detailed theoretical study of line profiles in
a mixed Hydrogen/Helium environment is required to examine
the effect on the ionization balance (see Bildsten 2000).

We also have to consider the possibility that the lines seen
in EXO 0748-676 are not gravitational
redshifted iron lines from the surface. EXO\,0748--676 is a high
inclination system and shows large dips in the X-ray light curve
due to an absorbing medium in the outer accretion disk, far
from the inner part of the system. Homan, Wijnands, \& van den
Berg (2003) proposed that a large fraction of the bursts used by
Cottam et al. (2002) occurred during times of significant
absorption at low energies ($<2$ keV) due to obscuring medium
in the outer disk. This might have introduced absorption features in the burst
spectra which could be due to the material inside the
obscuring matter and not due to material on the surface of the
neutron star. However, it is difficult to associate the lines seen in
the burst spectra of 
EXO\,0748--676 with He-like or H-like lines from abundant elements unless
large velocity shifts are invoked.  In contrast, if obscuration in the
outer disk drives dipping behavior, small velocity shifts would be
expected.  On balance, it is not clear that obscuration in the outer
disk could produce the absorption lines observed.  

Additional observations of burst sources with {\it XMM-Newton} can
help to better understand the conditions in which lines in X-ray
bursts are strongest.     Future X-ray missions with high collecting
area and high spectral resolution such as {\it Constellation-X} and
{\it XEUS} are ideally suited to detecting absorption lines from the
surface of neutron stars.  We ultimately look forward to definitive
observations with these missions.

\begin{acknowledgements}
This work is based on observations obtained with \xmm, an ESA mission
with instruments and contributions directly funded by ESA member
states and the US (NASA). W.H.G.L. gratefully acknowledges support from
the NASA.
\end{acknowledgements}

{\it Facilities:} \facility{XMM (EPIC, RGS)}

\end{document}